\documentclass[aps,prl,groupedaddress,fleqn,twocolumn]{revtex4}
\pdfoutput=1
\usepackage{graphicx}
\usepackage{amsmath}
\usepackage{gensymb}
\usepackage{multirow,array}
\usepackage{adjustbox}

\begin{document}
\title{Thermodynamic heterogeneity and crossover in the supercritical state of matter}
\author{L. Wang$^{1*}$}
\author{C. Yang$^{1,2}$}
\author{M. T. Dove$^{1}$}
\author{V. V. Brazhkin$^{3}$}
\author{K. Trachenko$^{1}$}
\address{$^1$ School of Physics and Astronomy, Queen Mary University of London, Mile End Road, London, E1 4NS, UK}
\address{$^2$ Institute of Natural Sciences, Shanghai Jiao Tong University, Shanghai 200240, China}
\address{$^3$ Institute for High Pressure Physics, RAS, 108840, Moscow, Russia}

\begin{abstract}
A hallmark of a thermodynamic phase transition is the qualitative change of system thermodynamic properties such as energy and heat capacity. On the other hand, no phase transition is thought to operate in the supercritical state of matter and, for this reason, it was believed that supercritical thermodynamic properties vary smoothly and without any qualitative changes. Here, we perform extensive molecular dynamics simulations in a wide temperature range and find that a deeply supercritical state is thermodynamically heterogeneous, as witnessed by different temperature dependence of energy, heat capacity and its derivatives at low and high temperature. The evidence comes from three different methods of analysis, two of which are model-independent. We propose a new definition of the relative width of the thermodynamic crossover and calculate it to be in the fairly narrow relative range of 13-20\%. On the basis of our results, we relate the crossover to the supercritical Frenkel line.
\end{abstract}

\maketitle

Transitions between different phases and properties of those transitions have been one of central themes of condensed matter physics. Ideas and methods developed in this field are applied to other wide-ranging areas, including geology, chemistry, quantum field theory and cosmology. First and second-order phase transitions and related critical phenomena are best-studied \cite{ma}. A first-order transition takes place across the boiling line. The line finishes at the critical point where the transition becomes second-order.

Phase transitions involve phases traditionally defined as physically uniform and homogeneous states or regions of space and which have qualitatively different properties such as different temperature or pressure dependencies of system characteristics. Interestingly, an absence of a phase transition in a certain range of thermodynamic parameters (pressure, temperature) does not necessarily imply that no qualitatively different states exist in that range. In other words, a transition between two states with qualitatively different properties can still take place even when the states belong to the same phase in the traditional definition of this term. When this occurs, we are dealing with a thermodynamic non-uniformity (heterogeneity) where a smeared transformation or a crossover takes place between the two states, rather than a phase transition. Compared to phase transitions, crossovers are less understood in general, not least because they are more subtle and are more difficult to detect, but also because their understanding is not developed from a general theoretical standpoint.

One example of what is believed to be a physically uniform and homogeneous state is the supercritical state of matter. Supercritical fluids are increasingly deployed in a number of important applications but their theoretical understanding is not developed \cite{deben}. The supercritical state is loosely defined to be above the range of thermodynamic parameters above the critical point. It was widely thought that no qualitative differences exist between gases and liquids, or any physical states, in the supercritical regime \cite{ma,deben}. This belief implied that thermodynamic properties vary smoothly and without any qualitative changes. Guided by a theoretical consideration of particle dynamics and modelling results \cite{prl,phystoday,pre}, we have recently proposed that two qualitatively different states with gas-like and liquid-like properties exist in the supercritical state and are separated by a crossover centered at the Frenkel line (FL). Recently, structural and dynamical changes at the FL have been experimentally confirmed in supercritical neon \cite{neon}, methane \cite{methane} and CO$_2$ \cite{co2}. Here, we focus on important {\it thermodynamic} properties of this crossover and discuss the evidence for the thermodynamic crossover seen as the change of central system properties such as energy, heat capacity and its derivatives.

We start by recalling the main idea of the Frenkel line in the supercritical state, based on the qualitative change of particle dynamics. At low temperature, the dynamics of liquid particles combine small-amplitude solid-like oscillatory motion around quasi-equilibrium positions and large-scale diffusive jumps between neighbouring positions, the picture introduced by Frenkel \cite{frenkel}. This motion is quantified by the liquid relaxation time $\tau$, the average time between particle's jumps. $\tau$ is directly related to viscosity and other important liquid properties. The FL proposal is based on considering how $\tau$ changes with temperature (pressure). Below the critical point, $\tau$ decreases with temperature until the liquid crosses the phase transition line and boils. Above the critical point where no phase transition intervenes, $\tau$ continuously decreases with temperature until it reaches its limiting value comparable with the shortest (Debye) transverse oscillation period $\tau_{\rm D}$. When $\tau\approx\tau_{\rm D}$, a particle spends about the same time oscillating as diffusing, at which point the oscillatory component of particle motion is lost. The crossover from combined oscillatory and diffusive particle motion to purely diffusive is the crossover at the FL. The FL is illustrated in Figure 1.

\begin{figure}
\begin{center}
{\scalebox{0.45}{\includegraphics{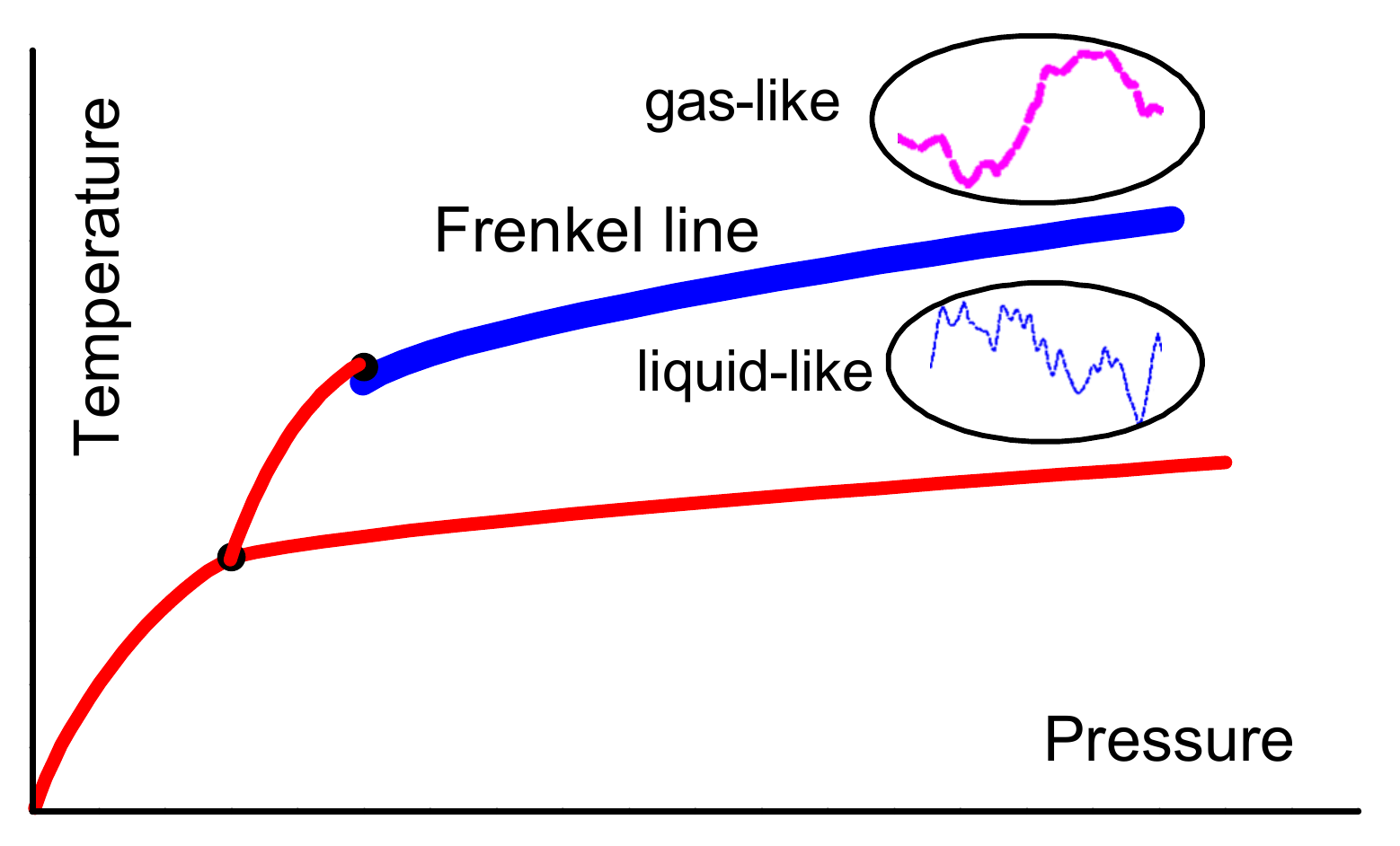}}}
\end{center}
\caption{The Frenkel line in the supercritical region. Particle dynamics includes both oscillatory and diffusive components below the line, and is purely diffusive above the line. Schematic illustration.}
\label{frenline}
\end{figure}

The dynamical crossover can be operationally defined from the disappearance of minima of velocity autocorrelation function (VAF), corresponding to monotonic VAF decay as in a gas \cite{prl}. This gives the FL on the phase diagram which coincides with $c_v=2$, where $c_v$ is specific heat and $k_{\rm B}=1$. There is a good reason for this coincidence because $c_v=2$ signifies a particular dynamical and thermodynamic state of the system as discussed below.

Frenkel predicted \cite{frenkel} that at short times $t<\tau$ or high frequency $\omega>\frac{1}{\tau}$ (by a factor of $2\pi$), the liquid behaves like a solid and therefore supports two solid-like transverse modes. The range in which transverse modes propagate shrinks with temperature, starting from the lowest frequency (or, to be more precise, from the smallest $k$-point or largest wavelength as discussed in detail later). Eventually, the liquid exhausts the frequency range at which it can support transverse modes, the result confirmed by direct molecular dynamics simulations \cite{jpcm}. At this point, liquid potential energy is given by the potential energy of one remaining longitudinal mode only, or $\frac{T}{2}$ per particle. Together with the kinetic contribution $\frac{3T}{2}$, the energy per particle becomes $2T$, giving the specific heat $c_v=2$ \cite{ropp}. On further temperature increase, it is now the longitudinal mode that starts shrinking in range, starting from the shortest wavelength because the wavelength can not be shorter than the particle mean free path \cite{ropp,natcom}.

Therefore, different collective modes evolve with temperature {\it differently} below and above the FL. This provides the key to predicting the thermodynamic crossover at the FL: since each mode contribute the energy $T$ per particle to the total energy of the supercritical system, the energy and its derivatives are predicted to have different temperature dependence below and above the FL (below we give specific equations for the system energy below and above the FL), i.e. exhibit a crossover around $c_v=2$. However, there has been no study of actual functional dependencies of the energy, $c_v$ and its derivatives below and above the FL and the associated crossover from its low-temperature to high-temperature regime. Another important open question is how wide is the thermodynamic crossover? Is the crossover range comparable to the entire range of variation of thermodynamic functions, or is it localised around the predicted $c_v=2$?

In this paper, we perform extensive molecular dynamics simulations of supercritical system in a very wide temperature range and analyze the functional dependence of energy, $c_v$ and its derivatives below and above the FL. Two of our analysis methods are model-independent and do not rely on any prior knowledge related to a crossover. The analysis shows that the supercritical state is thermodynamically heterogeneous, with different temperature dependence of energy, $c_v$ and its derivatives at low and high temperature. We find that the thermodynamic crossover corresponds to the Frenkel line, and is of a fairly narrow relative width of 13-20\%.

We have performed extensive parallel molecular dynamics (MD) simulations of supercritical liquid Ar using the Lennard-Jones (LJ) potential. This is a commonly-used potential which describes Ar with well-documented properties in a wide range of temperature and pressure (see, e.g. Ref. \cite{arg1}). We used the constant volume and energy ensemble with two densities, 1.5 and 1.9 g/ml, corresponding to system linear sizes of about 71 \AA\ and 65 \AA. The simulated temperature starts from just above the melting temperature and increases to very high temperatures deep in the supercritical state up to 50,000 K, corresponding to 215 times the critical temperature. In this work we study energy derivatives, $c_v$ and its derivative. This is often done by fitting the energy points and subsequently differentiating the fit. We find that the resulting $c_v$ and its derivatives can depend on the type of fit used. For this reason, we do not fit the energy. Instead, we simulate 510 temperature points for each density using a high-performance computing cluster. Each run involves 30 ps of equilibration and further 50 ps of production runs during which the calculated properties are averaged. Each temperature simulation is repeated 20 times using different starting conditions, and the calculated energy at each temperature is further averaged over 20 different runs. This averaging was found to be essential in order to reduce fluctuations of energy derivatives, particularly large at high temperature. In total, we performed 20,000 MD runs, equivalent to over 100,000 processor-hours or over 11 processor-years.

\begin{figure}
\begin{center}
{\scalebox{0.85}{\includegraphics{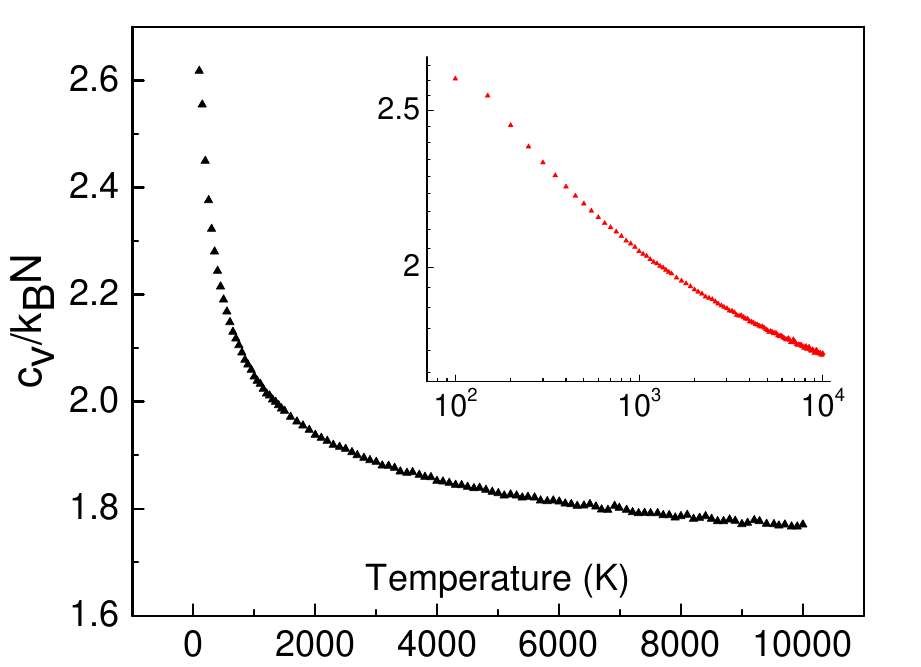}}}
{\scalebox{0.85}{\includegraphics{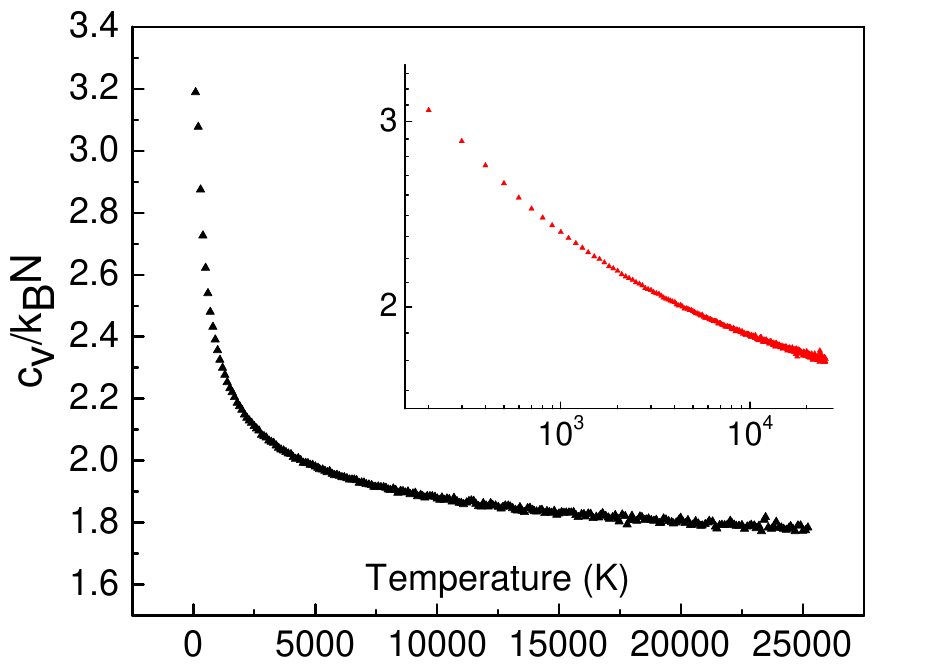}}}
\end{center}
\caption{$c_v$ calculated for density 1.5 g/ml (top) and 1.9 g/ml (bottom). The insets show the same data of each graph in the log-log plots for easily data reading.}
\label{cv}
\end{figure}

We focus on temperature dependence of energy $E$, $c_v=\frac{1}{N}\frac{dE}{dT}$ and its derivatives. We show $c_v$ at two densities in Figure \ref{cv}. As discussed above, the predicted crossover of thermodynamic properties takes place at temperature corresponding to $c_v=2$. This corresponds to the temperature range of about $1000-4000$ K at both densities. This temperature range corresponds to approximately $10-30T_c$ (Ar critical temperature is $T_c=151$ K) and hence the crossover can not be attributed to the Widom line, the line of persisting near-critical anomalies \cite{widom} which disappears after about 2$T_c$ \cite{widom1,widom2}.

We note that the crossover at $c_v=2$ assumes that each mode contributes $\frac{T}{2}$ to the energy as in the harmonic case. Intrinsic anharmonicity, present in experiments and MD simulations, can alter $c_v$ somewhat \cite{ropp}, implying that the crossover of $c_v$ takes place around $c_v=2$ but not exactly at it. 

We perform three methods of analysis. In the {\it first}, model-independent method, we focus on the temperature dependence of the system energy and fit the low-temperature and high-temperature parts of the energy to the simplest three-parameter function with a varying power exponent, $E=A+BT^n$. The high-temperature fitting range starts from temperature 3000 K above the predicted crossover range so that the analysis is not affected by the prior knowledge of the crossover temperature and finishes at 50,000 K, over ten times the crossover temperature. The low-temperature fitting range starts from just above the critical temperature and finishes 1000 K below the crossover range. The coefficient of determination related to the goodness of the fit, $R^2$, is significantly higher if the fit is performed for low- or high-temperature range as compared to the entire range: in the first case $1-R^2$ is $10-100$ times larger than in the second case. We define and calculate the difference parameter $D$:

\begin{equation}
D=(E-E_h)^2+(E-E_l)^2
\end{equation}

\noindent where $E_h$ and $E_l$ are fitted energies in the high- and low-temperature range, respectively.

$D$ can be used to study whether a given curve represents one or two different functional dependencies. If $E(T)$ is described by a single function, $D$ is zero or a small value related to fitting errors. If low and high-temperature dependence of $E(T)$ is given by two different functions, $E_l$ and $E_h$ will make zero contributions to $D$ at low and high $T$, respectively and, conversely, large contributions to $D$ at high and low $T$. This, together with $D$ being positive, implies that $D$ has a minimum. Large $D$ at either low or high $T$ indicates that this temperature is far away from the crossover between the two functions. At the minimum, neither $E_h$ nor $E_l$ fit the function as well as they do on the asymptotes but do not make a large contribution to $D$. Hence, the minimum of $D$ approximately corresponds to the crossover between the two functions.

\begin{figure}
\begin{center}
{\scalebox{0.75}{\includegraphics{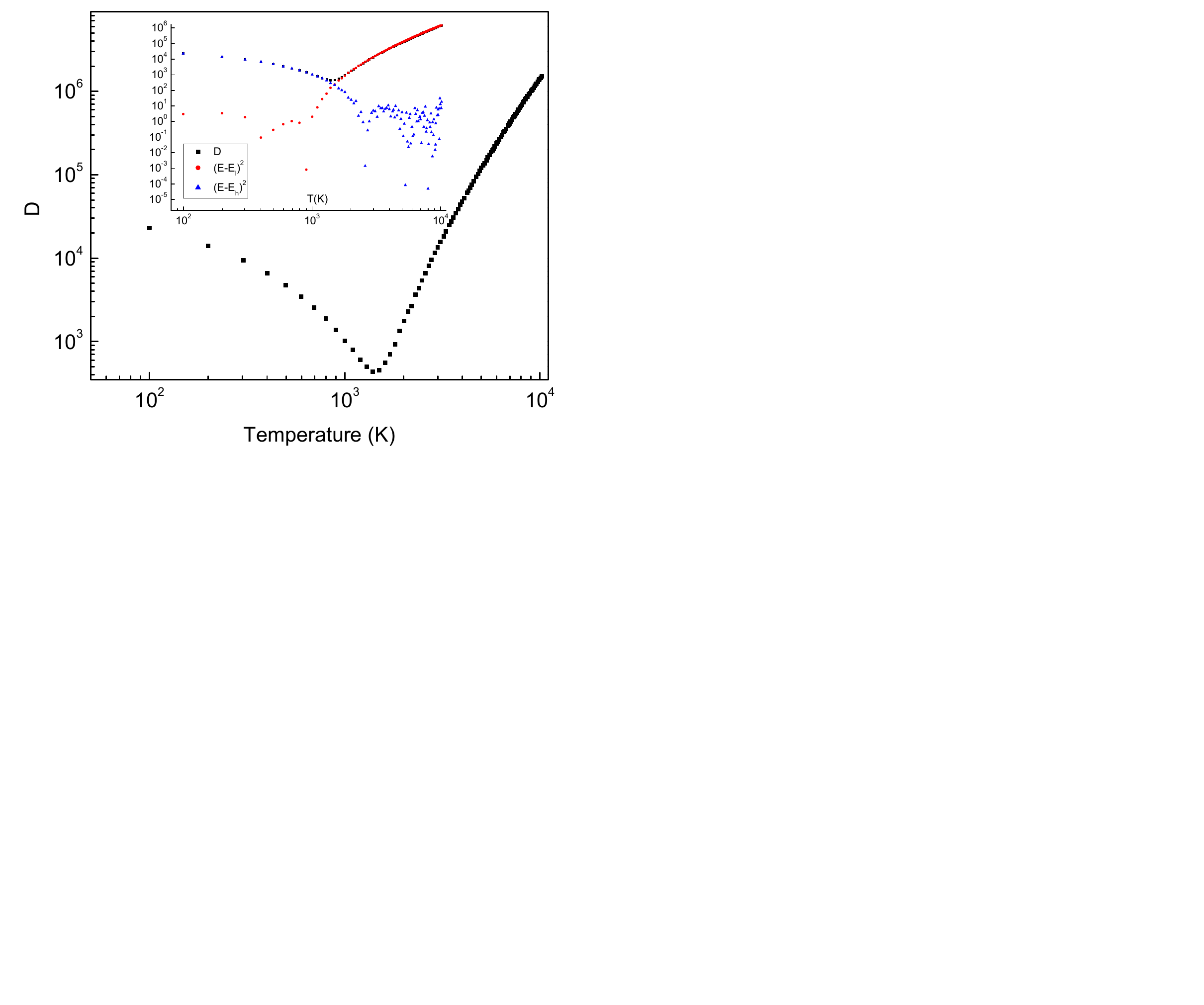}}}
{\scalebox{0.75}{\includegraphics{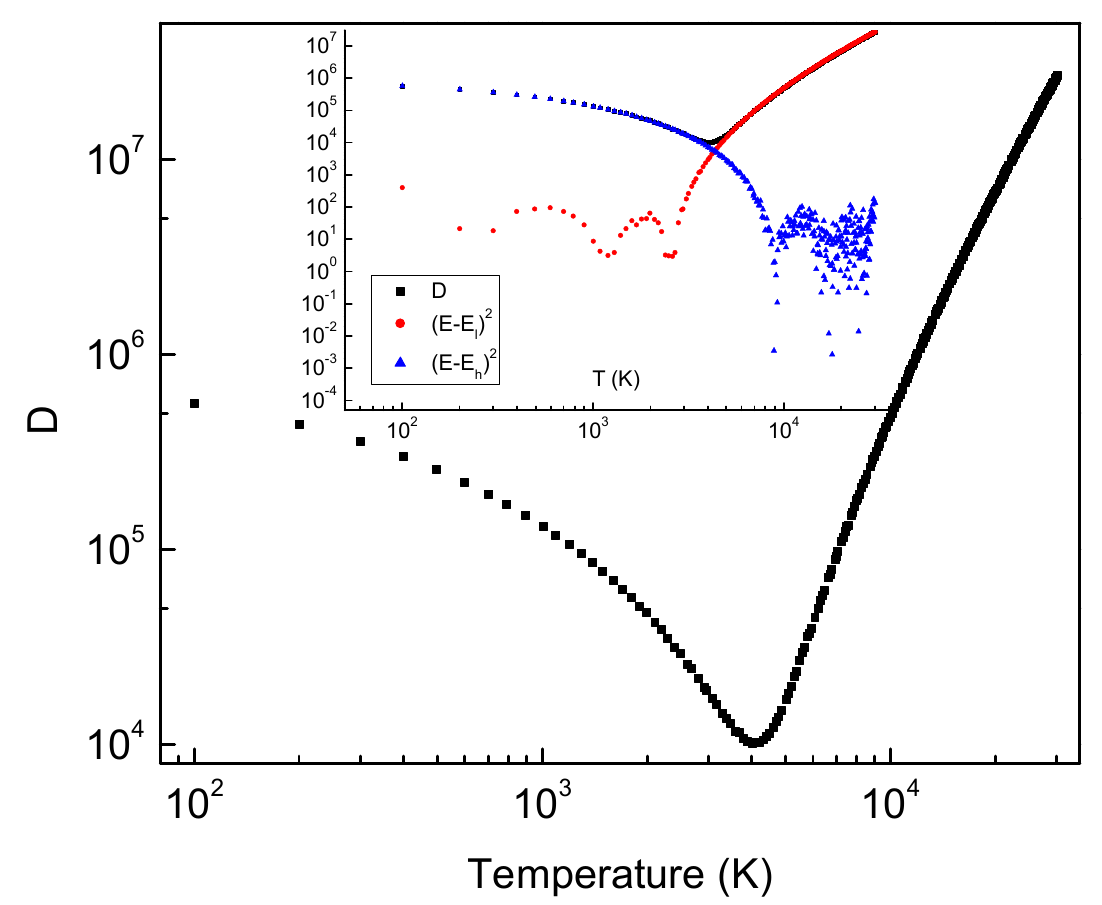}}}
\end{center}
\caption{Difference parameters $D$ calculated for density 1.5 g/ml (top) and 1.9 g/ml (bottom). The insets show $D$ and two terms of $D$: $(E-E_{l})^2$ and $(E-E_{h})^2$, respectively.}
\label{D}
\end{figure}

We plot $D$ for two densities in Figure \ref{D} and make two important observations. First, we observe a very deep minimum of $D$ for both densities. Second, the minimum develops at temperatures around 1000 K and 4000 K at both densities. We observe that these are the temperatures at which $c_v=2$ at both densities in Figure \ref{cv}, corresponding to the crossover at the Frenkel line.

Similarly to the first method, our {\it second} method of analysis is model-independent and deals with $c_v$ and its derivaties. We observe that $c_v$ changes slowly at high temperature in Figure \ref{cv} and crosses over to a much steeper function at low temperature. The double-logarithmic plot of $c_v$ is not linear (see the inset to Figure \ref{cv}), implying that the temperature dependence of $c_v$ can not be described by a power law. To address the change of slope of $c_v$ in more detail, we calculate $\frac{d c_v}{dT}$ using the data in Figure \ref{cv} and observe a slow and nearly constant behavior at high temperature which crosses over to a much steeper dependence at low temperature. The crossover from the small to large slope is in the range of about 1000-3000 K for both densities. This is close to the temperature range at which $c_v=2$ in Figure \ref{cv}, corresponding to the crossover at the FL.

\begin{figure*}
\begin{center}
{\scalebox{1.0}{\includegraphics{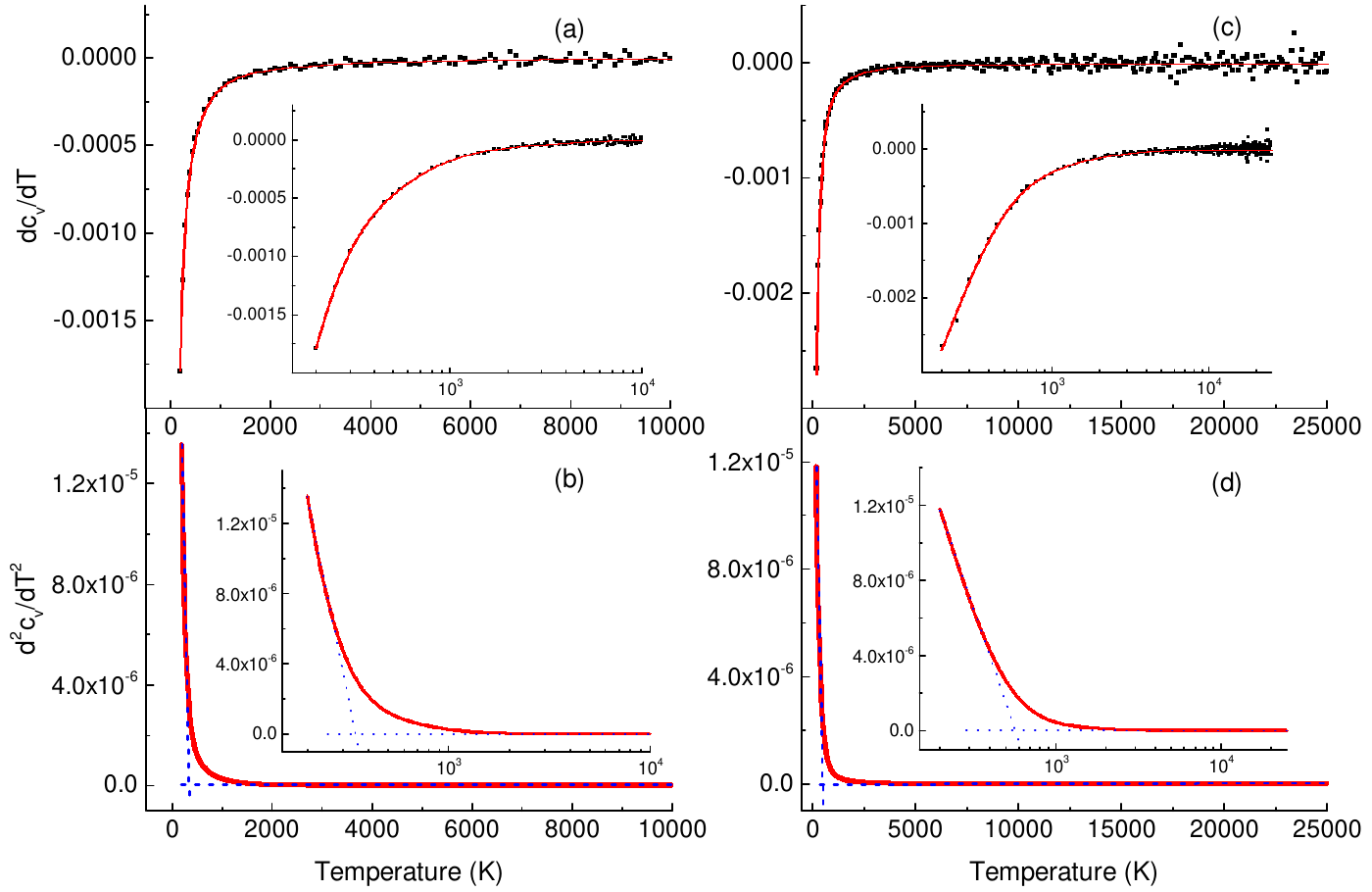}}}
\end{center}
\caption{(a) shows $\frac{d c_v}{dT}$ and its fit for density 1.5 g/ml; (b) shows $\frac{d^2 c_v}{dT^2}$ calculated as the derivative of the fit in (a), with the dashed lines showing straight-line approximations to the low- and high-temperature range. The insets show the same data of each graph in the semi-logarithmic plots to highlight the low-temperature range with steep slopes. (c) and (d) show the same for density 1.9 g/ml.}
\label{deriv}
\end{figure*}

Next, we fit $\frac{d c_v}{dT}$ and show the fit in Figure \ref{deriv}a (the inset shows the semi-logarithmic plot highlighting the low-temperature range with the large slope). Using the fit, we calculate the second derivative $\frac{d^2 c_v}{dT^2}$ and plot it in Figure \ref{deriv}b, together with the semi-logarithmic plot highlighting the low-temperature range with the steep slope). We observe a similar behavior of the second derivative: the crossover from slow high-temperature to steep low-temperature behavior. The crossover is in the range 1000-2000 K for both densities and is close to the temperature range where $c_v=2$ at the FL.

In view of the very wide temperature range considered in Figure \ref{cv}, the closeness of the crossover temperatures of first and second derivatives to temperatures where $c_v=2$ is particularly encouraging.

On the basis of Figure \ref{deriv}, we can discuss the width of the thermodynamic crossover, $W$. There is no universal approach to defining the crossover width. If low and high-temperature range of a function in question (or its certain transformation) are close to straight lines, the crossover width is given by the difference between temperatures at which deviations from the straight lines are seen. However, this definition is not unique and depends on the path chosen on the phase diagram (we have chosen the constant-volume path for convenience of calculating $c_v$). Indeed, a constant-pressure path gives a different $W$: density decrease with temperature results in faster temperature decrease of $\tau$ below the FL and faster increase of the mean free path $l$ above the FL and, therefore, gives a smaller $W$.

In order to alleviate the issues related to the dependence of the crossover width on the path on the phase diagram, we propose a definition of the crossover that is based on $c_v$ itself (or functions of $c_v$) rather than on thermodynamic parameters such as temperature or pressure. We define the relative width of the specific heat crossover as

\begin{equation}
W=\frac{c_v^l-c_v^h}{c_v^{cr}}
\label{cvcross}
\end{equation}

\noindent where $c_v^l$ and $c_v^h$ are values of $c_v$ corresponding to the deviation of $c_v$ or its function from their low- and high-temperature behavior and $c_v^{cr}=2$ is the value of $c_v$ at the FL crossover.

Observing that low- and high-temperature range of $\frac{d^2 c_v}{dT^2}$ can be approximated by the straight lines, we find $c_v^l$ and $c_v^h$ from taking the deviation temperatures from Figure \ref{deriv}b,d and subsequently finding corresponding $c_v$ from Figure \ref{cv}. We note that there is a slight ambiguity in finding the deviation temperatures related to how one chooses to draw the straight-line approximation, particularly in the low-temperature range. Whereas this may affect the deviation temperature, the corresponding variation of $c_v$ in Figure \ref{cv} is insignificant. We find $c_v^l=2.15$ and $c_v^h=1.895$ for low density and $c_v^l=2.3$ and $c_v^h=1.9$ for high density. Eq. (\ref{cvcross}) gives $W$ of 13\% and 20\% at low and high density, respectively. The smaller $W$ is close to the width of the experimental {\it structural} crossover of 10-12 \% in supercritical Ne \cite{neon}.

In our {\it third} method of analysis, we continue to analyze the temperature dependence of the system energy. This method is based on linearizing the energy by transforming the energy functions into linear dependencies using recent theoretical predictions. This method of analysis might be viewed as a less general analysis method, however it is widely used in the analysis of experimental and modelling data. The key to calculating the energy below the FL is accounting how the range of solid-like transverse modes shrinks with temperature. In the Frenkel picture where transverse modes propagate above frequency $\omega_{\rm F}=\frac{1}{\tau}$, the energy of transverse modes per particle is calculated to be \cite{ropp} $E_t=2T\left(1-\left(\frac{\omega_{\rm F}}{\omega_{\rm D}}\right)^3\right)$, where $\omega_{\rm D}$ is Debye frequency. A more detailed analysis \cite{ropp} gives the dispersion relation of transverse modes as $\omega=\sqrt{c^2k^2-\frac{1}{\tau^2}}$, which implies the gap in $k$-space, $k_g=\frac{1}{c\tau}$. Ascertained on the basis of direct modelling results \cite{prlgap}, the gap implies that the range of transverse modes shrinks with temperature because $\tau$ increases. $E_t$ can be calculated as either the integral over $k$-space or frequency, and gives the same $E_t$ as above \cite{prlgap,jpcm}. As shown previously \cite{ropp}, adding the energy of the remaining longitudinal mode and the kinetic energy gives the total liquid energy per particle as:

\begin{equation}
E=E_0+T\left(3-\left(\frac{\omega_{\rm F}}{\omega_{\rm D}}\right)^3\right)
\label{e1}
\end{equation}

\noindent where $E_0$ is the temperature-independent term giving the classical energy at zero temperature.

Using (\ref{e1}), the energy below the FL can be linearized as follows. Using the commonly considered Vogel-Fulcher-Tammann (VFT) law for the temperature dependence of $\omega_{\rm F}=\omega_{\rm D}e^{-\frac{U}{T-T_0}}$, where $U$ and $T_0$ are VFT parameters, we introduce $f_l$ as $f_l=\frac{1}{\ln\left(3-\frac{E-E_0}{T}\right)}$. According to (\ref{e1}), $f_l=-\frac{T-T_0}{3U}$. Hence, $f_l$ is predicted to be a linear function of temperature.

Calculating the energy above the FL involves considering how the remaining longitudinal mode changes with temperature. Recall that above the Frenkel line, the oscillatory component of particle motion is lost and only the gas-like diffusive motion remains. As temperature increases, the particle mean free path $l$ grows. Because the system can not oscillate at wavelengths shorter than $l$, $l$ sets the smallest wavelength of the remaining longitudinal wave in the system. As shown previously \cite{ropp,natcom}, adding the energy of this mode to the kinetic energy of atoms gives the total energy per particle of the supercritical fluid above the FL as:

\begin{equation}
E=E_0+\frac{3}{2}T+\frac{1}{2}T\left(1+\frac{\alpha T}{2}\right)\left(\frac{a}{l}\right)^3
\label{e2}
\end{equation}

\noindent where $a$ is the interatomic separation and $\alpha$ is the coefficient of thermal expansion that makes an anharmonic contribution to the mode energy, significant at high temperature above the FL.

Using (\ref{e2}), the energy above the FL can be linearized by introducing $f_h=\frac{E-E_0-\frac{3}{2}T}{1+\frac{1}{2}\alpha T}$. According to (\ref{e2}), $f_h=\frac{1}{2}T\left(\frac{a}{l}\right)^3$. $l$ can be calculated from the gas-like viscosity as $\eta=\frac{1}{3}\rho{\bar v}l$, where $\rho$ is density and ${\bar v}$ is average velocity. Using experimental $\eta$, we have earlier shown that $l$ depends on temperature as power law, $l\propto{T^\alpha}$ \cite{natcom}. The same result follows from the kinetic gas theory: approximating the Enskog series by the first term and considering the interatomic interaction in the form of the inverse-power law $U\propto\frac{1}{r^m}$ gives viscosity $\eta$ as $\eta\propto T^s$, where $s=\frac{1}{2}+\frac{2}{m-1}$ \cite{chapman}. Combining it with $\eta=\frac{1}{3}\rho{\bar v}l$, we find $l\propto T^{\frac{2}{m-1}}$ because ${\bar v}\propto T^{\frac{1}{2}}$. Hence, $f_h=\frac{1}{2}T\left(\frac{a}{l}\right)^3$ follows the power law and is the linear function in the double-logarithmic plot.

\begin{figure}
\begin{center}
{\scalebox{0.85}{\includegraphics{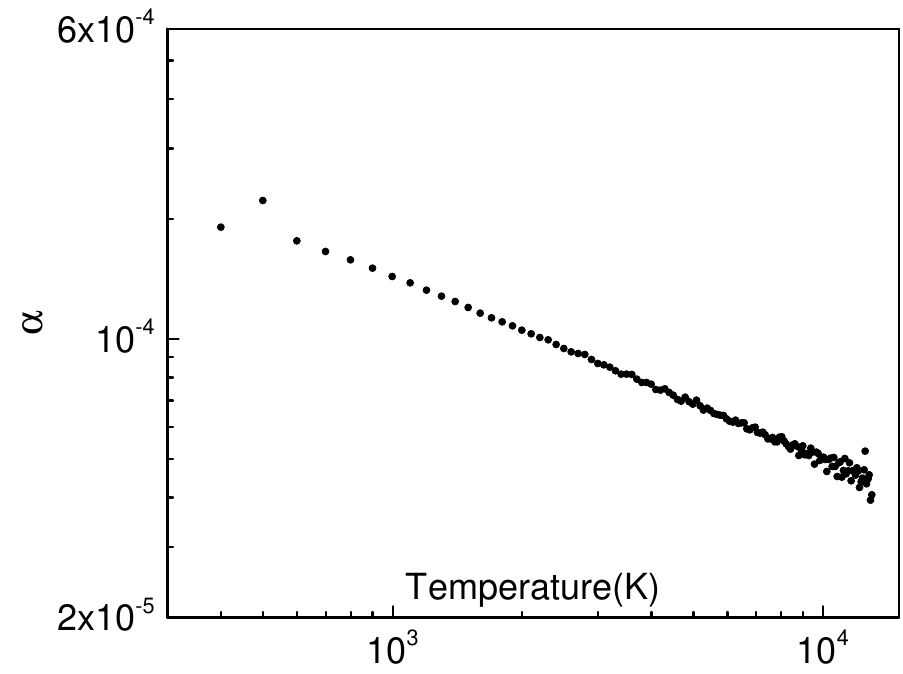}}}
\end{center}
\caption{Thermal expansion coefficient at density $\rho=$1.9 g/ml.}
\label{alpha}
\end{figure}

To calculate $f_h$, we need to first determine $\alpha$. Since our energy calculations are at constant density, we perform new calculations at constant pressure enabling the calculation of $\alpha=\frac{1}{V_0}\frac{V-V_0}{T}$. $\alpha$ depends on pressure which increases with temperature in the constant-density simulations used for energy calculations. To calculate $\alpha$, we used the pressure in the middle of the pressure range in constant-density simulations. The calculated $\alpha$ as a function of temperature is shown in Figure \ref{deriv}. Interestingly, we observe that $\alpha$ in the gas-like regime above the FL decreases approximately as inverse power law, as expected from the ideal gas equation of state. Using $\alpha(T)$ from Figure \ref{alpha}, we calculate $f_h$ at each temperature.

\begin{figure}
\begin{center}
{\scalebox{0.33}{\includegraphics{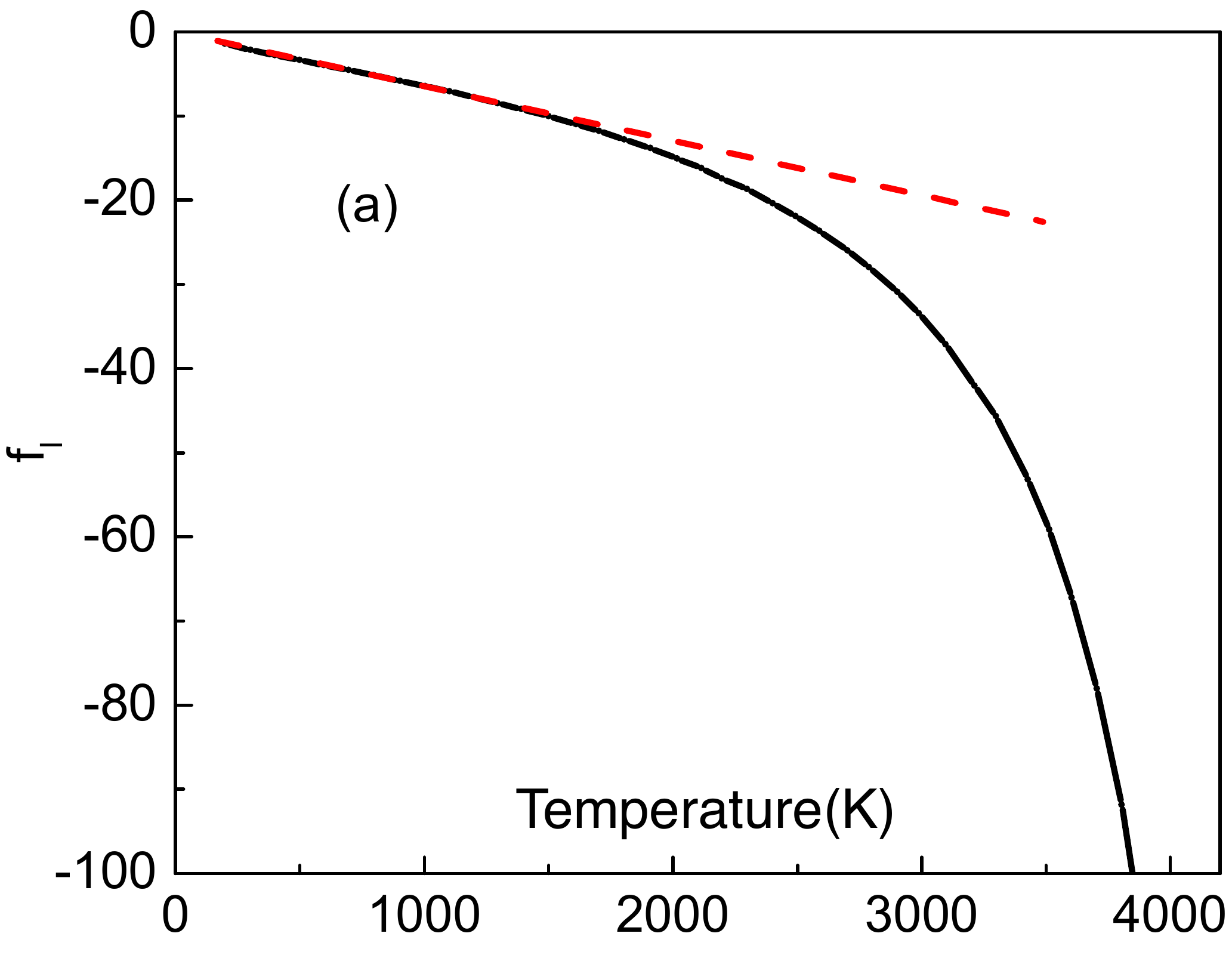}}}
{\scalebox{0.335}{\includegraphics{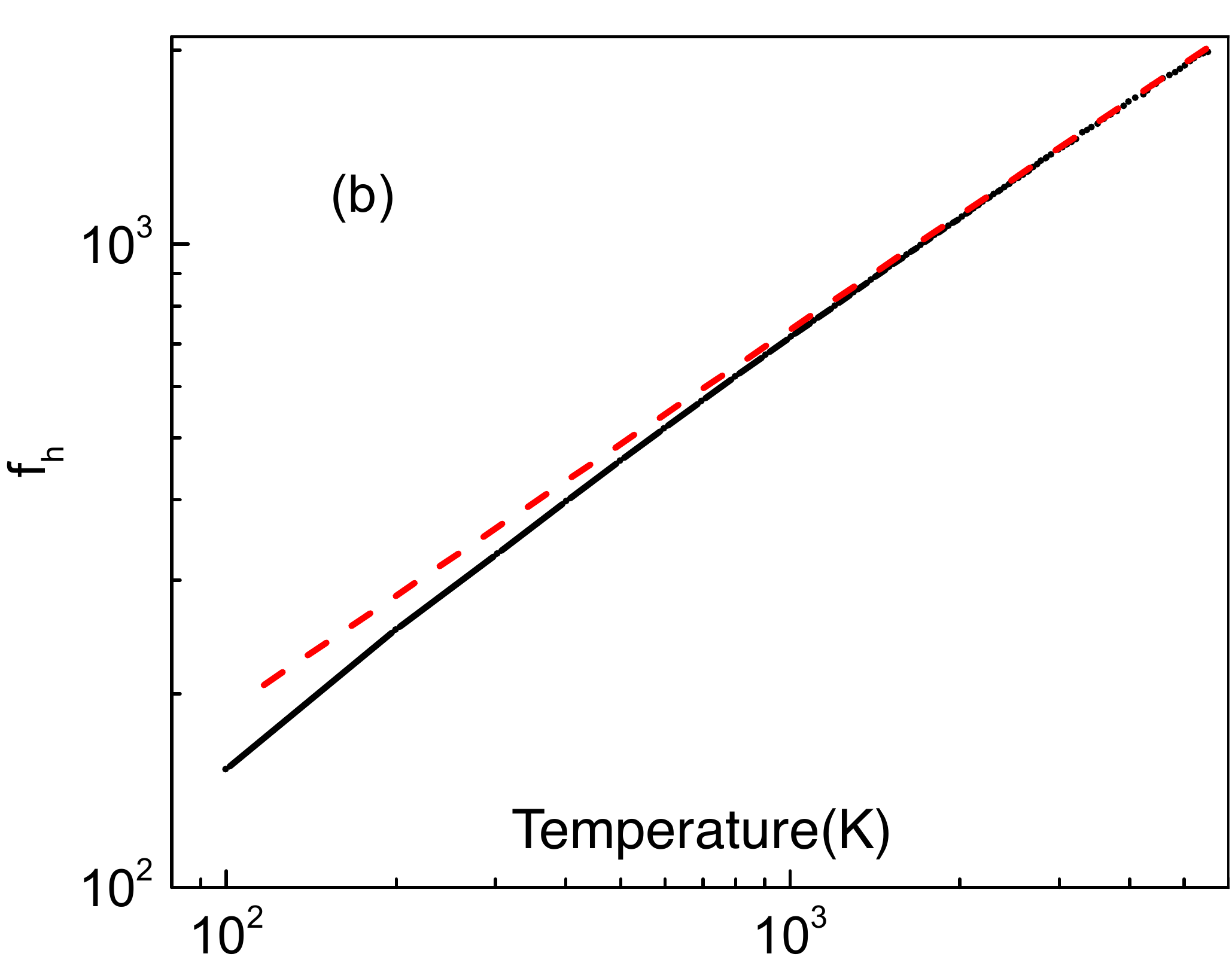}}}
{\scalebox{0.8}{\includegraphics{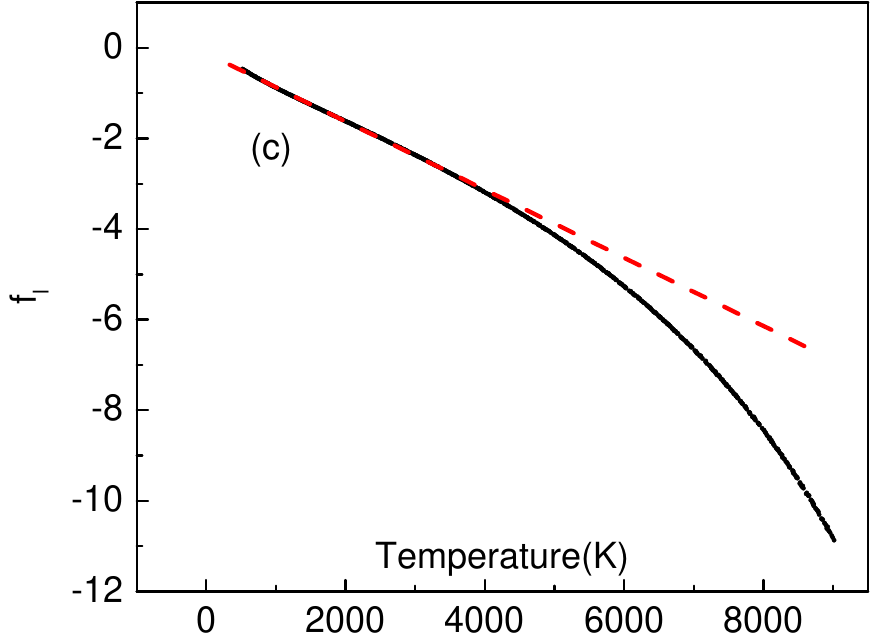}}}
{\scalebox{0.8}{\includegraphics{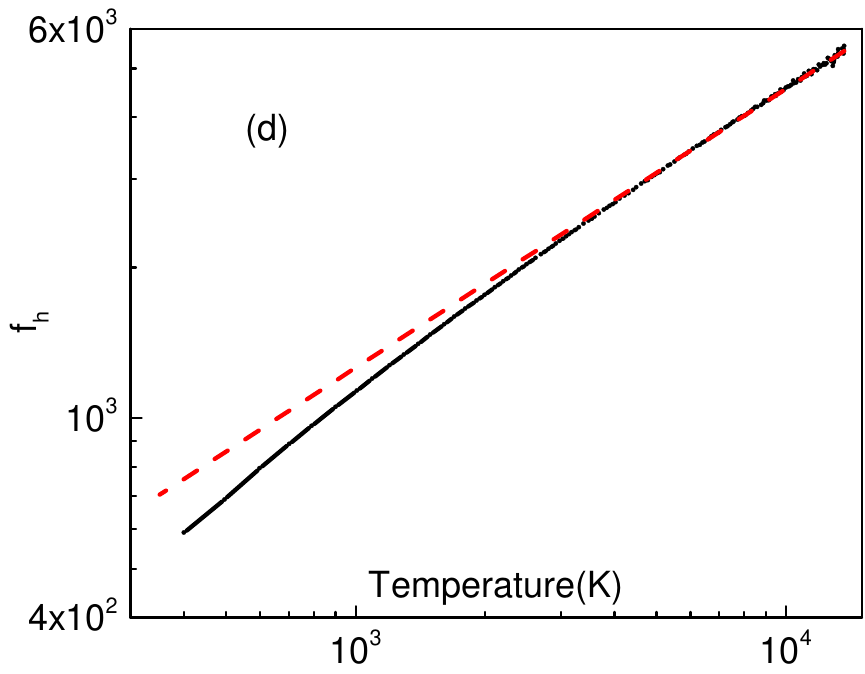}}}
\end{center}
\caption{$f_l$ and $f_h$ shown at low density in (a)-(b) and high density in (c)-(d). The dashed lines show the linear regimes at low and high temperature, respectively.}
\label{f}
\end{figure}

We plot $f_l$ and $f_h$ at two densities in Figure \ref{f}. Consistent with theoretical predictions, we observe a linear temperature dependence of $f_l$ in the low-temperature range and linear slope of $f_h$ in the high-temperature range in the double-logarithmic plot. Deviations from the linearity are seen and take place at temperatures close to the crossover temperature at both densities. This is expected because the theoretical results (\ref{e1}) and (\ref{e2}) are designed to work for $2<c_v<3$ and $1.5<c_v<2$, respectively. The important insight from this analysis is that linearizing the energy using a theoretical model works for either low- or high-temperature range but not both. This provides further support to the previous results that there are two supercritical regimes characterized by different temperature dependencies of the energy, with an accompanying thermodynamic crossover.

In addition to linearizing energy functions, Eqs. (\ref{e1}) and (\ref{e2}) provide a physical explanation for why $c_v$ has two different functional dependencies at low and high temperature and its very different slopes in particular. The physical properties of interest are $\omega_{\rm F}$ in (\ref{e1}) and $l$ in (\ref{e2}). $\omega_{\rm F}$ in (\ref{e1}) increases faster-than-exponential with temperature, namely as the VFT law discussed above. This gives fast disappearance of transverse modes with temperature and large slope of $c_v$ and its derivatives in Figure \ref{deriv}). On the other hand, $l$ in (\ref{e2}) is a slow function of temperature. Indeed, $l\propto T^\alpha$ with $\alpha$ close to 0.1, as followed from the experimental gas-like viscosity $\eta=\frac{1}{3}\rho{\bar v}l$ \cite{natcom}. The same result follows from the kinetic gas theory \cite{chapman} as mentioned earlier: $l\propto T^{\frac{2}{m-1}}$, where $m$ is the exponent of the interaction potential $U\propto\frac{1}{r^m}$. We recall that the function describing the repulsive part of the LJ potential is governed by both repulsive $\frac{1}{r^{12}}$ and attractive $\frac{1}{r^6}$ terms, with the net result that the effective $m$ is 18-20 \cite{dyre}. Consistent with the previous result, this gives $\alpha$ close to 0.1, explaining why $c_v$ and its derivative are slow functions of temperature.

Before concluding, we note that the although the highest temperatures simulated in this work might seem unusually high, there are three reasons why they are relevant to real systems and experiments. First, liquid argon remains an unmodified system up to fairly high temperature: the first ionization potential of condensed liquids is on the order of 10$^5$ K. Hence the simulated temperature range corresponds to the unmodified non-ionized argon describable by the LJ potential. Second, performing experiments at realistic constant pressure, rather than constant density used here, lowers the crossover temperature significantly due to faster increase of the mean free path when volume increases. Third, performing the experiments in systems with lower critical point such as Ne implies that the crossover temperature is lower: the crossover at the largest temperature simulated here is predicted to be lower by the ratio of Ar and Ne critical temperatures, or over 3 times.

In summary, we have established thermodynamic heterogeneity of the supercritical state using three different methods of analysis. The associated thermodynamic crossover has the relative width in the range 13-20 \% and corresponds to the Frenkel line. Our results do not preclude the existence of a phase transition in a narrow temperature range close to $c_v=2$ and call for a more detailed investigation of this point.

This research utilized Queen Mary's MidPlus computational facilities, supported by QMUL Research-IT and funded by EPSRC Grant No. EP/K000128/1. L.W., C.Y., M.T.D., and K.T. are grateful to the Royal Society and CSC. V.V.B. is grateful to the RSF (Grant No. 14-22-00093).

\end{document}